# A Deep Neural Network-based Frequency Predictor for Frequency-Constrained Optimal Power Flow

Fan Jiang
Department of Electrical and Computer Engineering
*University of Houston*
Houston, TX, USA
fjiang6@uh.edu

Xingpeng Li
Department of Electrical and Computer Engineering
*University of Houston*
Houston, TX, USA
xli83@central.uh.edu

Pascal Van Hentenryck
School of Industrial and Systems Engineering
*Georgia Institute of Technology*
Atlanta, GA, USA
pvh@gatech.edu

*Abstract*—Rate of change of frequency (RoCoF) and frequency nadir should be considered in real-time frequency-constrained optimal power flow (FCOPF) to ensure frequency stability of the modern power systems. Since calculating the frequency response is complex, deep neural network (DNN) could be adopted to capture the nonlinearities and estimate those two metrics accurately. Therefore, in this paper, a DNN-based frequency predictor is developed with the training data obtained from time-domain simulations using PSCAD/EMTDC. Subsequently, it is reformulated using a set of mixed-integer linear programming formulations and then embedded into the FCOPF framework as constraints to ensure grid frequency stability, creating the proposed DNN-FCOPF model. Two benchmark models, a traditional OPF without any frequency constraints and a linear system-wide RoCoF-constrained FCOPF, are also implemented to gauge the proposed DNN-FCOPF. Finally, the solutions obtained with these three models are compared and evaluated with time-domain simulations using PSCAD under various load profiles, demonstrating the effectiveness of the proposed DNN-FCOPF.

## I. INTRODUCTION

Frequency stability has become increasingly critical due to the emerging low inertia characteristics of modern power systems. To ensure the frequency response remains within an acceptable range, system-level methods like frequency-constrained optimal power flow (FCOPF) which incorporates frequency constraints into economic operations, are being widely studied [1]-[6]. Frequency response can be divided into three processes based on response time: initial response (IR), primary response (PR) and secondary response (SR). The SR, governed by automatic generation control (AGC), balances load and generator output imbalances, with a response time typically on the order of tens of seconds [7]. Therefore, in this paper, we focus primarily on the IR and PR for the FCOPF model. Two key metrics to prevent generator tripping during IR and PR are the rate of change of frequency (RoCoF) and frequency nadir [8].

The frequency transient response following a generator outage is complex, and accurate calculation of the RoCoF and frequency nadir requires solving a large set of differential and algebraic equations. However, FCOPF, as a real-time generation dispatch model, demands rapid computational performance. Consequently, these frequency constraints must be simplified when incorporated into FCOPF. Some studies assume the minimum value of RoCoF occurs immediately after the disturbance, and thus, the expression of RoCoF can be approximately obtained from the swing equation of generators [1]-[3], [7], which leads to a linear RoCoF constraint. For the frequency nadir constraints, which are affected by PR, the approaches can be divided into two categories: indirectly constrained by primary reserves of all generators [1], [6], [9] and directly constrained by an equation derived by simplifying the governor control model [4], [7]. However, these models are often not sufficiently accurate due to the simplifications they involve. In summary, the RoCoF and frequency nadir calculation methods are either too complex to solve or not accurate enough.

Machine learning (ML) offers a promising solution to address the above-mentioned issues. ML models can be trained offline using a large amount of data to improve its accuracy [10]. Then, the trained ML model can replace complex nonlinear constraints in optimization models such as unit commitment (UC) and optimal power flow (OPF). For OPF problems, some researchers have attempted to use ML to solve them directly [11]-[13], while others focus on replacing certain OPF constraints to improve the calculation speed [3], [8], [14]. The study shows that the end-to-end OPF direct solution using ML may cause infeasible solutions, while replacing part of the optimal model with ML models do not have that issue [12]. Therefore, replacing the frequency stability constraints with ML in FCOPF is a feasible method to improve both the solving speed and accuracy of the FCOPF problem.

Existing studies in replacing frequency constraints with ML are limited. Some studies applied this approach in UC problems. [3] proposes a deep neural network (DNN)-based frequency nadir predictor for UC and then ensure system stability through the encoding framework. However, these UC methods are not suitable for FCOPF, because the system inertia can be optimized by committing generators, which is not available in real-time situations. For OPF problems, a DNN-based inertia management framework for FCOPF is proposed in [4]. However, only the frequency nadir calculation is replaced by a DNN predictor, the RoCoF constraint still uses the simplified linear equation.

Therefore, a DNN-based frequency predictor including both RoCoF and frequency nadir for FCOPF is needed, which is proposed in this paper. Firstly, a DNN-based frequency predictor for RoCoF and frequency nadir is proposed, using electromagnetic transient (EMT) simulation results as training data to ensure its precision. Then, this predictor is incorporated into FCOPF, which creates the proposed DNN-FCOPF model, as a set of mixed-integer linear constraints to ensure grid frequency stability. The DNN frequency predictor is validated by comparing results with those of a real-time EMT model. Finally, a traditional OPF (T-OPF), the linear FCOPF (L-FCOPF) and the DNN-FCOPF are tested on the IEEE 9-bus system. Their performance is evaluated with against EMT simulations, and the results of frequency nadir and RoCoF between the three models demonstrate that the proposed DNN-FCOPF model can significantly improve both solving efficiency and accuracy.

## II. DEEP NEURAL NETWORK-BASED FREQUENCY PRECITOR

This section proposes the DNN-based frequency predictor for RoCoF and frequency nadir, along with the methodology for obtaining the training data.

### A. The DNN-based Frequency Predictor

DNN is a type of deep learning model, designed to mimic the neural structure of human brain for learning tasks. In this paper, it serves as the frequency predictor.

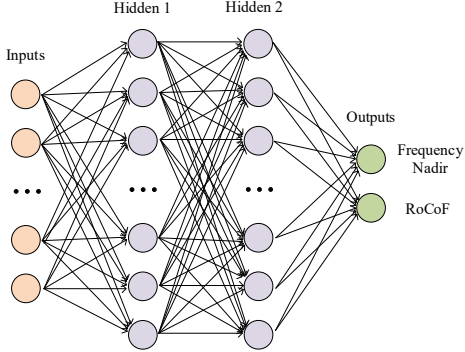

Fig. 1. Structure of the DNN-based frequency predictor.

Fig. 1 illustrates the structure of the DNN-based frequency predictor. Each neuron in the hidden layers and the output layer is fully connected to all neurons in the respective previous layers. The neurons in each layer are connected to the neurons of the previous layer through a polynomial transformation, and then are activated by an activation function, preparing for the next layer to be connected. In this paper, the inputs of the predictor are the real power of all generators, the real power of all loads and the location of the tripped generator. The outputs are frequency nadir and RoCoF. The equations of the predictor can be described by (1)-(4). Each layer except the output layer uses ReLU as the activation function, which is shown in (3).

$$z_1 = xW_1 + b_1 \quad (1)$$
$$z_m = a_{m-1}W_m + b_m, 2 \leq m \leq NL \quad (2)$$
$$a_m = \max(z_m, 0), \ 1 \leq m \leq NL \quad (3)$$
$$R_h = a_{NL}W_{NL+1} + b_{NL+1} \quad (4)$$

where, $W_m$, $b_m$ represent weights and biases of layer $m$, $NL$ denotes the number of hidden layers, $a_m$ is the value of neurons of layer $m$, $x$ stands for the inputs, and $R_h$ is the output vector.

### B. Training Data Acquisition Approach

Obtaining a high-quality dataset is essential for training the frequency predictor and enhancing its accuracy, as the dataset quality significantly impacts the performance of the predictor. Most studies acquire the frequency response data after generator tripping using PSS/E, a software mostly focuses on the electromechanical rather than EMT phenomena. However, obtaining the accurate frequency response metrics especially RoCoF requires an EMT simulation model. PSCAD/EMTDC, a well-established commercial platform, is widely used for power system frequency transient studies. Therefore, in this paper, a model developed in PSCAD/EMTDC is utilized to generate the dataset, ensuring it accurately captures EMT dynamics.

The training dataset is obtained through a large number of independent runs of the simulation model developed in PSCAD, considering different loads, initial generators outputs, as well as the locations of the tripped generator. The RoCoF and frequency nadir are calculated during each simulation. The measuring window of the RoCoF typically ranges from 5 cycles to 10 cycles [15]. In this paper, we use 10 cycles, 0.167s for 60Hz system. Worst measured values are used as the RoCoF labels. It is important to note that the frequency responses vary across different buses. Frequency performance will be worse on buses closer to the disturbance point. Therefore, the frequency metrics should be measured at the bus where the disturbance occurs.

After acquiring the RoCoF and frequency nadir of all scenarios, the dataset are then used for the training process. The trained frequency predictor is then incorporated into the FCOPF model as the frequency constraints.

## III. FREQUENCY-CONSTRAINED OPTIMAL POWER FLOW

This section explains the objectives and constraints of (i) a traditional T-OPF without any frequency constraints, (ii) a linear system-wide RoCoF-constrained L-FCOPF, and (iii) the proposed DNN-FCOPF with a DNN-based frequency predictor.

### A. Traditional optimal power flow model

In this paper, the quadratic functions are used as the generator cost functions, then the objective of the T-OPF is:

$$C(P_{gi}) = \sum c_{2i}P_{gi}^2 + c_{1i}P_{gi} + c_{0i} \quad (5)$$

where, $i$ is the generator index, $P_g$ stands for the generator real power output, and $c_2$, $c_1$, $c_0$ are the generator cost coefficients.

The nodal power balance equations, the power flow equations, the generator outputs constraints, and the transmission lines thermal limitation constraints are shown in (6)-(9) respectively.

$$\sum P_{k,from} + P_{load} = \sum P_{k,to} + P_g \quad (6)$$
$$P_k = (\theta_{k,from} - \theta_{k,to})/x_k \quad (7)$$
$$P_g^{min} \leq P_g \leq P_g^{max} \quad (8)$$
$$P_k^{min} \leq P_k \leq P_k^{max} \quad (9)$$

where, $k$ is the transmission line index, $\theta$ represents the phase angle, and $x_k$ stands for the line reactance.

### B. Linearization of RoCoF and frequency nadir

Most Independent System Operators (ISOs) of power systems run OPF repeatedly in real-time around every 5 minutes, which requires a good computational speed of the model. Besides, the system frequency response involves higher-order inertial elements, and calculating the higher-order equations is quite time-consuming. Therefore, it is necessary to linearize the frequency response equations in the FCOPF model.

Generally, the worst RoCoF (hereinafter referred to as RoCoF) is assumed to occur at the beginning of the generator outage [1]-[3], [7]. So, the RoCoF can be acquired by the swing equation of generator. The swing equation is shown in (10).

$$\Delta P_m - \Delta P_e = \frac{2H_{sys}P_{base}}{f_0}\frac{df}{dt} \quad (10)$$

where, $\Delta P_m$ and $\Delta P_e$ stand for the mechanical input and electrical output power change respectively, $H_{sys}$ is the equivalent system inertia, $P_{base}$ is the real power base value, and $f_0$ is the frequency base value.

For a short period of time after the disturbance, the mechanical input power $\Delta P_m$ is assumed to remain unchanged, while the $\Delta P_e$ changes immediately. Therefore, the RoCoF is:

$$RoCoF = -\frac{f_0}{2H_{sys}P_{base}}P_{loss} \quad (11)$$

where $P_{loss}$ is the lost generator's output before tripping.

Then, the RoCoF constraint is as follows:

$$-\frac{f_0}{2H_{sys}P_{base}}P_{loss} \geq RoCoF_{threshold} \quad (12)$$

The frequency nadir is constrained by the primary reserves in many papers [1], [6], [9], or it is directly obtained by using a simplified governor response model in other papers. Both methods have nonlinear part in their equations, and they are time-consuming. Therefore, frequency nadir constraints are not included in the L-FCOPF model.

In summary, the L-FCOPF model is an enhanced T-OPF with an extra constraint of (12).

*C. OPF model with DNN-based frequency predictor*

In the DNN-based frequency predictor, the activation function is nonlinear. Generally, it can be linearized by (13)-(16).

$$a_m \leq z_m - h_l(1 - B_m) \quad (13)$$
$$a_m \geq z_m \quad (14)$$
$$a_m \leq h_u B_m \quad (15)$$
$$a_m \geq 0 \quad (16)$$

where $h_l$ and $h_u$ are the lower boundary and upper boundary of the value of all possible $z_m$, and $B_m$ is a binary variable.

The outputs of the frequency predictor are frequency nadir ($f_{nadir}^{DNN}$) and RoCoF ($RoCoF^{DNN}$) as shown in Fig. 1, then the constraints about frequency in the DNN-FCOPF model are:

$$RoCoF^{DNN} \geq RoCoF_{threshold} \quad (17)$$
$$f_{nadir}^{DNN} \geq f_{threshold} \quad (18)$$

Therefore, the DNN-FCOPF is enhancing T-OPF with additional constraints (13)-(18).

To summarize, the objective and constraints of the three OPF models are arranged in TABLE I.

TABLE I. Objectives and constraints of T-OPF, L-FCOPF and DNN-FCOPF

| Model | Objective | Shared Constraints | Unique Constraints |
|---|---|---|---|
| T-OPF | | | None |
| L-FCOPF | (5) | (6)-(9) | (12) |
| DNN-FCOPF | | | (13)-(18) |

## IV. CASE STUDIES

This section first implements the three OPF models, then validates the accuracy of the proposed DNN-based frequency predictor, and finally presents a comparative analysis of the T-OPF, L-FCOPF, and DNN-FCOPF models based on their performance on a modified IEEE 9-bus system.

The topology of the modified IEEE 9-bus system is illustrated in Fig. 2. Generators on bus 1, 2 and 3 are split into multiple different generators: two at bus 1, four at bus 2, and three at bus 3. This aims to maintain the contingency capacity between 10% and 15%, preventing the severe disturbances which is infrequent in real power systems. It is notable that the generators at the same bus are identical in all aspects, including rated capacity, governor response control and cost function, so they have the same outputs.

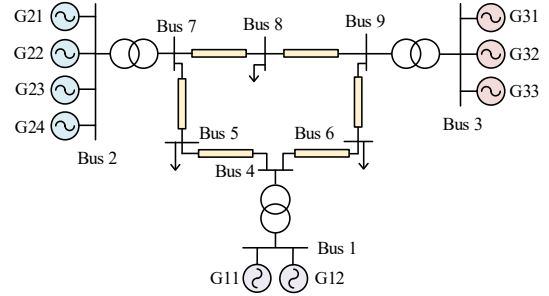

Fig. 2. Illustration of a modified IEEE 9-bus system.

*A. Development of T-OPF, L-FCOPF and DNN-FCOPF*

In the default modified IEEE 9-bus system, the sum of generator outputs at bus 1, 2 and 3 are 72MW, 163MW and 85MW, and each generator output can be calculated by dividing the number of generators at this bus. The real power of the original loads at bus 5, 6 and 8 are: 125MW, 90MW, 100MW. Based on this system, T-OPF, L-FCOPF and DNN-FCOPF are implemented according to the functions in Section III. All the settings are the same for these three models, except the frequency constraints. The following provides a detailed procedure for establishing the proposed DNN-FCOPF model.

For the development of the DNN-based frequency predictor, the dataset must first be obtained from a reliable source. To generate this dataset and verify the performance of the proposed DNN-FCOPF model, a real-time simulation model is developed in PSCAD/EMTDC. Governor BPA_GG is used as the governor control, while the exciter model is IEEE AC5C. To meet the *N*-1 requirement, the tripping of G11, G21, or G31 is considered as the contingencies. The loads in this model vary from 90% to 110% of original load values, and generator outputs are approximately from 85% to 115% in each scenario. Then, the dataset is acquired through this time-domain EMT model. Each data point consists of the RoCoF and frequency nadir for a given scenario, along with its corresponding conditions, including loads, generator outputs, and contingency locations. This model can also serve as a reference for validating the frequency metrics of the FCOPF.

After acquiring the dataset, the DNN-based frequency predictor is trained and validated (details provided in the next section), then linearized by (13)-(16) and embedded into the proposed DNN-FCOPF model in Python 3.9, using Gurobi as the solver. The DNN-FCOPF is solved on a computer equipped with a 12th Gen Intel® Core™ i7-12700 CPU @ 2.1 GHz and 32.0 GB of RAM.

*B. Validation of DNN-based frequency predictor*

The accuracy of DNN-FCOPF is mostly determined by the DNN-based frequency predictor, so it is important to train a well-performing model. The training process uses mean square error (MSE) as the loss function, which is defined in (19).

$$MSE = \frac{1}{n}\sum(R_h - R_{true})^2 \quad (19)$$

where $n$ is the number of data points, $R_{true}$ stands for true value of each data point.

To validate the proposed DNN frequency predictor, the loads at bus 5, 6 and 8 assumed to be adjusted to 90% of the original load values due to the weather changes or other factors. The changes in load are balanced by generators at bus 1 (slack bus). Considering the tripping of G11, the respective errors of frequency nadir and RoCoF are 0.016% and 0.40% as summarized in TABLE II, which is quite low and acceptable for frequency metrics.

TABLE II. Results of EMT simulations and DNN-based frequency predictor

|  | EMT Simulation | DNN-based Frequency Predictor | Error |
|---|---|---|---|
| RoCoF | -0.495 Hz/s | -0.497 Hz/s | 0.016% |
| $f_{nadir}$ | 59.64 Hz | 59.63 Hz | 0.40% |

Therefore, the proposed DNN frequency predictor offers a high level of accuracy in forecasting both frequency nadir and RoCoF, and is suitable to be incorporated into OPF models.

*C. Comparison of T-OPF, L-FCOPF and DNN-FCOPF*

This subsection compares the results of T-OPF, L-FCOPF and DNN-FCOPF models. Meanwhile, the modified IEEE 9-bus system developed in PSCAD is used to verified the results of the three OPF models.

According to the description in Section IV.A, the generator dispatch results of the three OPF models are used to set the generator outputs of the simulation model developed in PSCAD. The contingency is assumed to be the tripping of G21 at 0s, which is the most critical contingencies.

Under the same conditions, the $f_{threshold}$ and $RoCoF_{threshold}$ in the L-FCOPF and DNN-FCOPF models are set to 59.5Hz and -0.5Hz/s respectively [2]. The results of the three optimization OPF models and the corresponding EMT simulation results are summarized in TABLE III. The calculation of the errors of RoCoF and $f_{nadir}$ are shown in (20). This function uses the results from the simulation model developed in PSCAD as the reference, where the initial outputs of the generators are initialized based on the dispatch results of the corresponding OPF model.

$$Error = \frac{(y_{OPF} - y_{PSCAD})}{y_{PSCAD}} \times 100\% \quad (20)$$

where $y_{OPF}$ is the result from OPF model and $y_{PSCAD}$ represents the result from the corresponding EMT simulation.

As shown in TABLE III, the frequency metrics in DNN-FCOPF have the lowest errors under all contingencies, while the errors of L-FCOPF are pretty high. This indicates that the proposed DNN-FCOFP has a better performance on the frequency security.

TABLE III. Results of T-OPF, L-FCOPF and DNN-FCOPF with G21 outage

|  | Model | T-OPF | L-FCOPF | DNN-FCOPF |
|---|---|---|---|---|
| OPF | Total Cost ($) | 5409.8 | 5409.8 | 5412.5 |
|  | Solve Time (s) | 0.27 | 0.36 | 2.07 |
|  | RoCoF (Hz/s) (% error) | N/A | -0.36 (28.9%) | -0.50 (1.2%) |
|  | $f_{nadir}$ (Hz) (% error) | N/A | N/A | 59.67 (0.02%) |
| PSCAD | RoCoF (Hz/s) | -0.831 | -0.831 | -0.506 |
|  | $f_{nadir}$ (Hz) | 59.46 | 59.46 | 59.66 |

The frequency response curves and associated RoCoF curves in EMT simulations under this contingency are depicted in Fig. 3. It is notable that since the dispatch results of the T-OPF and L-FCOPF models are identical, their frequency response curves are also identical. The total generator outputs at bus 1, 2 and 3 from the DNN-FCOPF results are 45.79MW (22.89MW per generator), 187.44MW (46.86MW per generator), and 86.40MW (28.80MW per generator). The $f_{nadir}$ is 59.64Hz and the RoCoF is -0.5Hz/s. The values of frequency response metrics of this EMT simulation model developed in PSCAD are 59.62Hz and -0.536Hz/s, and the errors are 0.02% and 3.3% respectively. However, in the L-FCOPF model, the error of RoCoF reaches as high as 21.6%. These low errors of the DNN-FCOPF model proves its effectiveness in ensuring frequency security. Similarly, the results of DNN-FCOPF under other contingencies are comparable to those of the related simulation model.

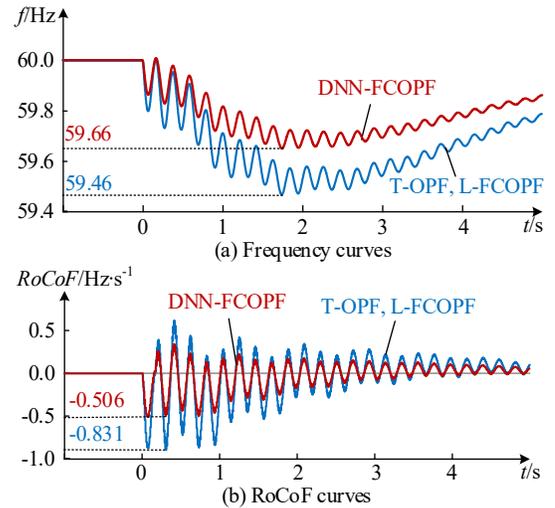

Fig. 3. (a) Frequency response curves and (b) RoCoF curves under G21 outage, using PSCAD, with initial grid conditions from T-OPF, L-FCOPF and DNN-FCOPF models respectively.

*D. High-Load Scenario Analysis*

To further validate the effectiveness of the proposed model, assume the loads increase to 120%, which means the loads at bus 5, 6, and 8 are 150MW, 108MW and 120MW. The contingency event is G11 outage at 0s. The comparison of the results of the three OPF models and related simulation models developed in PSCAD are summarized in TABLE IV. Additionally, the frequency response curves of the simulation models corresponding to the three OPF models are depicted in Fig. 4.

TABLE IV. Results of T-OPF, L-FCOPF and DNN-FCOPF with G11 outage in a high-load grid scenario

| | Model | T-OPF | L-FCOPF | DNN-FCOPF |
|---|---|---|---|---|
| OPF | Total Cost ($) | 5881.45 | 5881.48 | 5883.07 |
| | Solve Time (s) | 0.23 | 0.24 | 1.01 |
| | RoCoF (Hz/s) (% error) | N/A | -0.50 (45.11%) | -0.50 (7.07%) |
| | $f_{nadir}$ (Hz) (% error) | N/A | N/A | 59.65 (0.05%) |
| PSCAD | RoCoF (Hz/s) | -0.965 | -0.911 | -0.467 |
| | $f_{nadir}$ (Hz) | 59.32 | 59.36 | 59.68 |

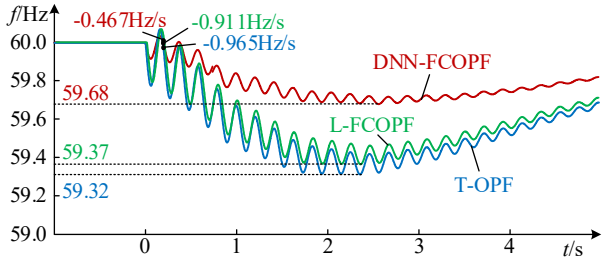

Fig. 4. Frequency response curves under G11 outage in a high-load grid scenario, using PSCAD, with initial grid conditions from T-OPF, L-FCOPF and DNN-FCOPF models respectively

It shows that the frequency response under the DNN-FCOPF dispatch results is much better than those of both T-OPF and L-FCOPF models.

In this scenario, the dispatch results of L-FCOPF differ from T-OPF. The reported RoCoF value of 0.5Hz/s for L-FCOPF, equaling to the threshold, indicates that the linearized frequency constraint is binding. However, the RoCoF error of this L-FCOPF model is as high as 45.11% comparing to the exact EMT simulation results. In contrast, the results in DNN-FCOPF are similar to the EMT simulations, with errors of RoCoF and frequency nadir being 7.07% and 0.05%, which verifying the accuracy of the proposed model. In addition, the load values fall outside the range of the training dataset for the DNN-based frequency predictor, which verifies the proposed model can perform well in out-of-distribution samples.

In summary, the comparisons of the T-OPF, L-FCOPF and DNN-FCOPF models under different loads show that the proposed DNN-FCOPF model is more realistic and can improve the accuracy of the FCOPF, ensuring grid frequency stability.

## V. CONCLUSION

This paper proposed a DNN-FCOPF model to enhance both the accuracy and solving efficiency of the FCOPF, improving the frequency reliability of the power system. The main contributions of this paper are summarized as follows:
(1) Proposed a DNN-based frequency predictor, whose training data are obtained from time-domain EMT simulations using PSCAD, which is more reliable.
(2) Embedded the predictor into OPF that leads to the proposed DNN-FCOPF model, which can balance the trade-off between the accuracy and solving time.
(3) Validated the efficiency and accuracy of the proposed DNN frequency predictor. Comparisons demonstrated the proposed DNN-FCOPF outperforms the two benchmark models T-OPF and L-FCOPF.

Built upon this work, the proposed DNN-FCOPF model will be extended to larger systems incorporating renewable energy with grid-following and grid-forming control strategies, which will be explored in our future research.

ACKNOWLEDGMENT

This research was (in part) supported by the National Science Foundation (NSF) under Grant No. 2337598; and it was partly supported by the NSF AI Institute for Advances in Optimization (Award 2112533) and the Georgia Tech Strategic Energy Institute. Any opinions, findings, and conclusions or recommendations expressed in this paper are those of the author(s) and do not necessarily reflect the views of the NSF.

REFERENCES

[1] K. Li et al., "Frequency security constrained robust unit commitment for sufficient deployment of diversified frequency support resources," IEEE Transactions on Industry Applications, vol. 60, no. 1, pp. 1725–1737, Jan. 2024.
[2] M. Tuo and X. Li, "Security-Constrained Unit Commitment Considering Locational Frequency Stability in Low-Inertia Power Grids," IEEE Transactions on Power Systems, vol. 38, no. 5, pp. 4134–4147, Sep. 2023.
[3] Y. Zhang et al., "Encoding Frequency Constraints in Preventive Unit Commitment Using Deep Learning With Region-of-Interest Active Sampling," IEEE Transactions on Power Systems, vol. 37, no. 3, pp. 1942–1955, May 2022.
[4] B. She, F. Li, H. Cui, J. Wang, Q. Zhang, and R. Bo, "Virtual Inertia Scheduling (VIS) for Real-Time Economic Dispatch of IBR-Penetrated Power Systems," IEEE Transactions on Sustainable Energy, vol. 15, no. 2, pp. 938–951, Apr. 2024.
[5] X. Zhao, H. Wei, J. Qi, P. Li, and X. Bai, "Frequency Stability Constrained Optimal Power Flow Incorporating Differential Algebraic Equations of Governor Dynamics," IEEE Transactions on Power Systems, vol. 36, no. 3, pp. 1666–1676, May 2021.
[6] H. Chávez, R. Baldick, and S. Sharma, "Governor Rate-Constrained OPF for Primary Frequency Control Adequacy," IEEE Transactions on Power Systems, vol. 29, no. 3, pp. 1473–1480, May 2014.
[7] H. Ahmadi and H. Ghasemi, "Security-Constrained Unit Commitment With Linearized System Frequency Limit Constraints," IEEE Transactions on Power Systems, vol. 29, no. 4, pp. 1536–1545, Jul. 2014.
[8] Q. Shi, F. Li, and H. Cui, "Analytical Method to Aggregate Multi-Machine SFR Model With Applications in Power System Dynamic Studies," IEEE Transactions on Power Systems, vol. 33, no. 6, pp. 6355–6367, Nov. 2018.
[9] G. Zhang and J. McCalley, " Optimal power flow with primary and secondary frequency constraint," in 2014 North American Power Symposium (NAPS), Sep. 2014, pp. 1-6.
[10] S. Qiu, Y. Li, Z. Wang, Y. Shen, Z. Li and F. Shen, "Geometric Matrix Completion for Missing Data Estimation in Power Distribution Systems," 2024 The 9th International Conference on Power and Renewable Energy (ICPRE), Guangzhou, China, 2024, pp. 1605-1609.
[11] J. Rahman, C. Feng, and J. Zhang, " Machine Learning-Aided Security Constrained Optimal Power Flow," in 2020 IEEE Power & Energy Society General Meeting (PESGM), Aug. 2020, pp. 1–5.
[12] C. Li et al., "Optimal Power Flow in a highly renewable power system based on attention neural networks," Applied Energy, vol. 359, p. 122779, Apr. 2024.
[13] T. Pham and X. Li, " Reduced Optimal Power Flow Using Graph Neural Network," in 2022 North American Power Symposium (NAPS), Oct. 2022, pp. 1-6.
[14] M. Tuo and X. Li, "Deep Learning based Security-Constrained Unit Commitment Considering Locational Frequency Stability in Low-Inertia Power Systems," Aug. 16, 2022, arXiv: arXiv:2208.08028. Available: http://arxiv.org/abs/2208.08028
[15] T. Amraee, M. G. Darebaghi, A. Soroudi, and A. Keane, "Probabilistic under frequency load shedding considering RoCoF relays of distributed generators," IEEE Trans. Power Syst., vol. 33, no. 4, pp. 3587–3598, Jul. 2018.